\documentstyle[prd,aps,twocolumn,psfig]{revtex}

 \newcommand{\CL}{{\cal L}}

 \newcommand{\bea}{\begin{eqnarray}}  \newcommand{\eea}{\end{eqnarray}}
 \newcommand{\beq}{\begin{equation}}  \newcommand{\eeq}{\end{equation}}
 \newcommand{\non}{\nonumber}  
   
 \newcommand{\lmk}{\left(}  \newcommand{\rmk}{\right)}

 \newcommand{\del}{\partial}  
 
 \newcommand{\bib}{\bibitem} \newcommand{\new}{\newblock}

\def\IB#1#2#3{{\bf #1}, #2 (19#3)}

\def\IBID#1#2#3{{\it ibid}. {\bf #1}, #2 (19#3)}
\def\IBIDD#1#2#3{{\it ibid}. {\bf #1}, #2 (20#3)}

\def\APJL#1#2#3{Astrophys. J. Lett. {\bf #1}, L#2 (19#3)}
\def\APJLL#1#2#3{Astrophys. J. Lett. {\bf #1}, L#2 (20#3)}

\def\IJMPD#1#2#3{Int. J. Mod. Phys. D {\bf #1}, #2 (19#3)}

\def\JP#1#2#3{J. Phys. A {\bf #1}, #2 (19#3)}

\def\NATT#1#2#3{Nature (London) {\bf #1}, #2 (20#3)}
\def\NPB#1#2#3{Nucl. Phys. {\bf B#1}, #2 (19#3)}

\def\PLB#1#2#3{Phys. Lett. B {\bf #1}, #2 (19#3)}

\def\PRD#1#2#3{Phys. Rev. D {\bf #1}, #2 (19#3)}
\def\PRDD#1#2#3{Phys. Rev. D {\bf #1}, #2 (20#3)}

\def\PRL#1#2#3{Phys. Rev. Lett. {\bf#1}, #2 (19#3)}

\def\PTP#1#2#3{Prog. Theor. Phys. {\bf #1}, #2 (19#3)}

\begin{document}
\twocolumn[\hsize\textwidth\columnwidth\hsize\csname
@twocolumnfalse\endcsname
\draft
\title{Cosmological evolution of global monopoles}
\author{Masahide Yamaguchi}
\address{Research Center for the Early Universe, University of Tokyo,
Tokyo, 113-0033, Japan}
\date{\today}
\maketitle

\begin{abstract}
    We investigate the cosmological evolution of global monopoles in
    the radiation dominated (RD) and matter dominated (MD) universes
    by numerically solving field equations of scalar fields.  It is
    shown that the global monopole network relaxes into the scaling
    regime, unlike the gauge monopole network. The number density of
    global monopoles is given by $n(t) \simeq (0.43\pm0.07) / t^{3}$
    during the RD era and $n(t) \simeq (0.25\pm0.05) / t^{3}$ during
    the MD era. Thus, we have confirmed that density fluctuations
    produced by global monopoles become scale invariant and are given
    by $\delta \rho \sim 7.2(5.0) \sigma^{2} / t^{2} $ during the RD
    (MD) era, where $\sigma$ is the breaking scale of the symmetry.
\end{abstract}

]

Since Kibble pointed out that topological defects may have been formed
as a consequence of cosmological phase transitions \cite{KIB}, many
cosmological applications have been investigated \cite{TD}. In
particular, the possibility is discussed that topological defects
should become a source to produce density fluctuations responsible for
the large-scale structure formation and the cosmic microwave
background (CMB) anisotropy \cite{TD}. Recently, the BOOMERANG
experiment \cite{BOOMERANG} and the MAXIMA experiment \cite{MAXIMA} on
observations of the CMB anisotropy revealed that our universe is flat,
which is consistent with the prediction of inflation. However, they
also found a relatively low second acoustic peak. One explanation of
such a feature is a hybrid model, where primordial fluctuations are
comprised of adiabatic fluctuations induced by inflation and
isocurvature ones by topological defects \cite{TDCMB}. In fact,
topological defects are compatible with the inflationary scenario
\cite{Yokoyama}. Moreover, deviations from Gaussianity in CMB are
reported in \cite{nonG}. Thus, it is still important to clarify the
dynamics of topological defects and their implications on cosmology.

Among many types of topological defects, particular attention has been
paid to strings. This is mainly because it is confirmed that both the
local\footnote{There are still discussions on the energy loss
mechanism of local strings, gravitational wave radiation via the loop
formation, or direct heavy particle emission \cite{VHS}.} \cite{local}
and the global string networks \cite{global,global2} relax into the
scaling regime, where the number of infinite strings per horizon
volume is a constant irrespective of cosmic time, so that density
fluctuations produced by strings become scale invariant. Thus, strings
have been examined as a possible source to produce density
fluctuations responsible for the large-scale structure formation and
the CMB anisotropy \cite{TD}. Furthermore, global strings have been
discussed in the context of axion cosmology
\cite{global,axion,axion2}. As the result of the breakdown of the
global Peccei-Quinn (PQ) $U(1)$ symmetry, global strings (axionic
strings) are formed. Then, axions emitted from axionic strings may
significantly contribute to the present energy density so that they
may play the role of the cold dark matter (CDM).

On the other hand, monopoles have been less investigated because the
formation of local monopoles, generally speaking, leads to conflict
with observation. If they are diluted by inflation \cite{inflation},
annihilated by the Langacker-Pi mechanism \cite{LP}, or swept away by
the domain wall \cite{DLV}, the problem can be evaded but local
monopoles leave no imprint on cosmological evolution.  Unlike strings,
however, the evolution of monopoles has distinct features depending on
whether they are local or global. In fact, the long range force works
between global monopoles so that their annihilations are much more
efficient than the local counterpart.  Therefore, the global monopole
network may go into the scaling regime, where the number of global
monopoles per horizon volume is a constant irrespective of cosmic time
\cite{BV,BR,PST}.

Bennett and Rhie made the first numerical simulations and found the
tendency that the number of global monopoles per horizon volume is a
constant though they used the nonlinear $\sigma$ model equation to
evolve the scalar fields \cite{BR}. Later, Pen, Spergel, and Turok
performed similar numerical simulations in both the nonlinear $\sigma$
model approximation and the full potential \cite{PST}. (See also
\cite{DKM}.) However, due to a lack of computer power, they can only
run a few realizations so that the number of global monopoles per
horizon volume still has large errors and the scaling property cannot
be definitely confirmed. Also, as will be shown later, we must pay
attention to the boundary effects, the grid size effects, and the
total box size dependence. In fact, the monopole can disappear under
the periodic boundary condition due to the boundary effects.

Now that we have sufficient computer power, we can solve the equation
of motion of the scalar fields $\phi^{a}$ without any approximations
and run many realizations in order to investigate and remove the
boundary effects, the grid size effects, and the total box size
dependence so that we can ascertain whether the global monopole
network enters the scaling regime. If so, the number density of global
monopoles is completely determined. In this paper, we report the
result of our numerical simulations for the evolution of the global
monopole network. Also, a simple analytic estimate is given.

As mentioned above, we directly solve the equation of motion for
scalar fields in the expanding universe. Let us consider the following
Lagrangian density for scalar fields $\phi^{a}(x)~(a = 1, 2, 3)$,
\beq
  \CL[\phi^{a}] = \frac12 g_{\mu\nu}\del^{\mu}\phi^{a}\del^{\nu}\phi^{a}
                 - V_{\rm eff}[\phi^{a},T],
\eeq 
where $g_{\mu\nu}$ is the flat Robertson-Walker metric and
the effective potential $V_{\rm eff}[\phi^{a},T]$ is given by
\bea
  V_{\rm eff}[\phi^{a},T] 
       &=& \frac{\lambda}{4}(\phi^{2} - \sigma^2)^2 
                 + \frac{5}{24}\lambda T^2 \phi^{2}, \non \\
       &=& \frac{\lambda}{4}(\phi^{2} - \eta^2)^2 
            + \frac{\lambda}{4}(\sigma^{4} - \eta^4). 
  \label{eqn:effpot}
\eea
Here $\phi \equiv \sqrt{\phi^{a}\phi^{a}}$, $\eta \equiv \sigma
\sqrt{1-(T/T_{c})^2}$, and $T_{c} \equiv \frac25\sqrt{15}\sigma$ is the
critical temperature. For $T > T_{c}$, the potential $V_{\rm
eff}[\phi^{a},T]$ has a minimum at the origin and the $O(3)$ symmetry
is restored. On the other hand, for $T < T_{c}$, new minima $\phi =
\eta$ appear and the symmetry is broken. In this case the phase
transition is of second order.

Then, the equation of motion for $\phi^{a}$ in the expanding universe
is given by
\bea
  \lefteqn{ \ddot{\phi^{a}}(x) + 3H\dot{\phi^{a}}(x) -
     \frac{1}{R(t)^2} \nabla^2 \phi^{a}(x)} \non \\
  && \hspace{3.0cm} 
   + \lambda[\phi^{2}(x) - \eta^{2}]\phi^{a}(x) 
    = 0,
  \label{eqn:master}
\eea
where the dot represents time derivative and $R(t)$ is the cosmic
scale factor. The Hubble parameter $H = \dot R(t)/R(t)$ and the cosmic
time $t$ are given by
\bea
  H^2 &=& \frac{4\pi^{3}}{45 m_{\rm pl}^2} g_{*} T^4,
   ~~~
  t = \frac{1}{2H} \equiv \frac{\epsilon_{RD}}{T^2} 
    ~~~~~~~~({\rm for~RD}), \non \\
  H^2 &=& \alpha(T) \frac{4\pi^{3}}{45 m_{\rm pl}^2} g_{*} T^4,
   ~~
  t = \frac{2}{3H} \equiv \frac{\epsilon_{MD}}{T^{3/2}} 
    ~~({\rm for~MD}),  
  \label{eqn:hubble}
\eea
where $m_{\rm pl} = 1.2 \times 10^{19}$ GeV is the Planck mass, and
$g_{*}$ is the total number of degrees of freedom for the relativistic
particles. For the MD case, we have defined $\alpha(T)$ [$\alpha(T) >
1$] as $\alpha(T) \equiv \rho_{\rm mat}(T) / \rho_{\rm rad}(T) =
\alpha_c (T_{c}/T)$, where $\rho_{\rm mat}(T)$ is the contribution to
the energy density from nonrelativistic particles, $\rho_{\rm
rad}(T)$ the contribution from relativistic particles at the
temperature $T$, and $\alpha_{c} \equiv \rho_{\rm mat}(T_{c}) /
\rho_{\rm rad}(T_{c})$. We also define the dimensionless parameter
$\zeta$ as
\bea
  \zeta_{RD} &\equiv& \frac{\epsilon_{RD}}{\sigma}  =
     \lmk \frac{45}{16\pi^3g_{*}} \rmk^{1/2}
     \frac{m_{\rm pl}}{\sigma} 
    ~~~~~~~({\rm for~RD}),  \non \\
  \zeta_{MD} &\equiv& \frac{\epsilon_{MD}}{\sigma^{1/2}}  =
     \lmk \frac{5\sqrt{15}}{6\alpha_{c}\pi^3g_{*}} \rmk^{1/2}
     \frac{m_{\rm pl}}{\sigma}
    ~~~~({\rm for~MD}).  
  \label{eqn:zeta}
\eea
In our simulation, we take $\zeta_{RD,MD} = 10$ and $5$ to investigate
$\zeta$ dependence on the result.

We start the simulations at the temperature $T_{i} = 2T_{c}$, which
corresponds to $t_{i} = t_{c}/4$ (RD) and $t_{i} =
t_{c}/(2\sqrt{2})$ (MD). Since the $O(3)$ symmetry is restored at the
initial time ($T_{i} > T_{c}$), we adopt as the initial condition the
thermal equilibrium state with the mass,
\beq
  m = \sqrt{\frac{5}{12}\lambda(T_{i}^{2} - T_{c}^{2})},
\eeq
which is the inverse curvature of the potential at the origin at $t =
t_{i}$.

Hereafter we normalize the scalar field in units of $t_{i}^{-1}$, $t$
and $x$ in units of $t_{i}$, $\lambda$ is set to be $\lambda = 0.25$,
and the scale factor $R(t)$ is normalized as $R(1) = 1$.
 
We perform numerical simulations in seven different sets of lattice
sizes and spacings for the RD case and the MD case (See Tables
\ref{tab:set1} and \ref{tab:set2}.). In all cases, the time step is
taken as $\delta t = 0.01$. In the typical case (1), the box size is
nearly equal to horizon volume $(H^{-1})^{3}$ and the lattice spacing
to a typical core size of a monopole $\delta x \sim
1.0/(\sqrt{\lambda}\sigma)$ at the final time $t_{f}$. Furthermore, in
order to investigate the dependence of $\zeta$, we arrange case (7)
with $\zeta = 5$. We have simulated the system from 10~[(2), (3), (5)
and (6)] or 50~[(1), (4), and (7)] different thermal initial
conditions. Also, in order to investigate the effect of the boundary
condition (BC), we adopt the periodic BC and the reflective BC
[$\nabla^{2}\phi^{a}(x) = 0$ on the boundary].

In order to judge whether the global monopole network relaxes into the
scaling regime, we give time development of $\xi$, which is defined as
\beq
   \xi \equiv n(t) t^3,
\eeq
where $n(t)$ is the number density of global monopoles. Before
counting the number of global monopoles in the simulation box, we must
identify global monopoles. We introduce two identification methods. In
one method (I), we use a static spherically symmetric solution, which
is obtained by solving the equation
\beq
  \frac{d^2 \phi}{dr^2} + \frac{2}{r}\frac{d\phi}{dr}
   - 2\frac{\phi}{r^2} - \frac{dV_{\rm eff}[\phi,T]}{d\phi} = 0,
\eeq
where $\phi^{a}(r, \theta, \varphi) \equiv \phi(r)x^{a} / r$ with
$x^{1} = r\sin\theta\cos\varphi, x^{2} = r\sin\theta\sin\varphi, x^{3}
= r\cos\theta$, and the winding number $n = 1$. The boundary
conditions are given by
\bea
  \phi(r) &\rightarrow& \eta, \qquad (r \rightarrow \infty), 
     \non \\
  \phi(0) &=& 0.
\eea
Since spacetime is discretized in our simulations, a point with $\phi
= 0$ corresponding to a monopole core is not necessarily situated at a
lattice point. In the worst case, a point with $\phi = 0$ lies at the
center of a cube. We require that a lattice is identified with a
monopole core if the potential energy density there is larger than
that corresponding to the field value of a static spherically
symmetric solution at $r = \sqrt{3}\delta x_{\rm{phys}}/2$ [$\delta
x_{\rm{phys}} = R(t)\delta x$]. Moreover, in order to reduce the
error, we regard the identified lattices which are connected as one
monopole core. In the other method (II), a cubic box is identified to
include a monopole if all $\phi^{a} = 0$ ($a$ = 1, 2, 3) surfaces pass
through the cubic box. In this method, we also regard the identified
boxes which are connected as one monopole core. In fact, as shown
later, the results with these two identification methods coincide very
well.

\begin{figure}
  \begin{center}
    \leavevmode\psfig{figure=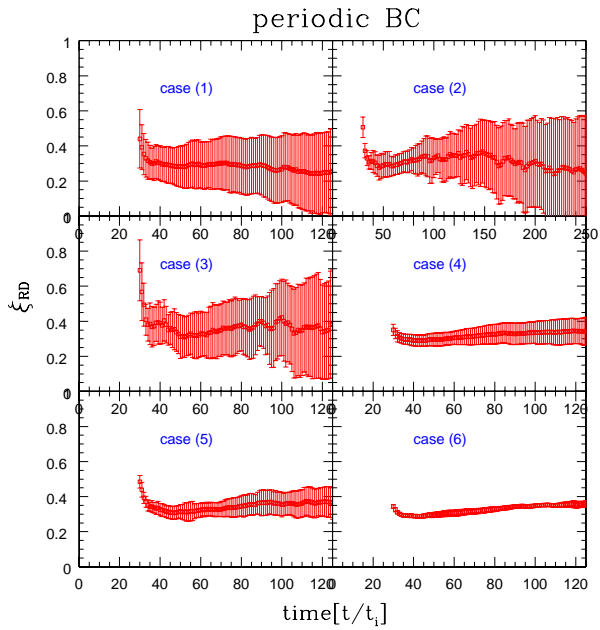,width=8.5cm}
  \end{center}
  \caption{Time development of $\xi_{RD}$ in cases (1) to
  (6) under the periodic BC for the RD case. Symbols ($\Box$) represent
  time development of $\xi_{RD}$. The vertical lines denote a standard
  deviation over different initial conditions.}
  \label{fig:xi}
\end{figure}

\begin{figure}
  \begin{center}
    \leavevmode\psfig{figure=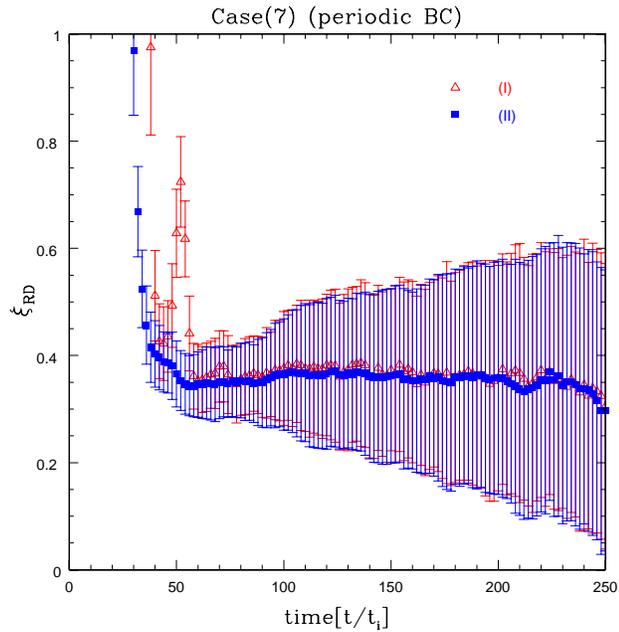,width=8.5cm}
  \end{center}
  \caption{Time development of $\xi_{RD}$ in case (7) under the
  periodic BC for the RD case. Filled squares represent time
  development of $\xi_{RD}$ for identification method (I) using the
  spherical symmetric solution, blank triangles for that (II) using
  the $\phi^{a} = 0$ surface. The vertical lines denote a standard
  deviation over different initial conditions.}
  \label{fig:xi2}
\end{figure}

As an example, time development of $\xi_{RD}$ in cases (1) to (6)
under the periodic BC for the RD case is described in Fig.
\ref{fig:xi}. We find that after some relaxation period, $\xi_{RD}$
becomes a constant irrespective of time for all cases. Though all are
consistent within the standard deviation, $\xi_{RD}$ tends to increase
as the box size does. This is because monopoles annihilate more often
than those in the real universe under the periodic BC so that
monopoles annihilate too much for smaller box sizes. On the other
hand, under the reflective BC, we also find that $\xi_{RD}$ becomes a
constant irrespective of time after some relaxation period. $\xi_{RD}$
is listed in Table \ref{tab:set1}. In this case, contrary to the case
under the periodic BC, $\xi_{RD}$ tends to decrease as the box size
increases.  This is because a monopole suffers a repulsive force from
the boundary so that monopoles near the boundary annihilate less often
and $\xi_{RD}$ takes larger values in smaller-box simulations due to
the boundary effect. Therefore, the real number of the monopole per
horizon volume lies in between those under the periodic and the
reflective BC. From the results of the largest-box simulations [case
(6)], we conclude that $\xi_{RD}$ converges to a constant $\xi_{RD}
\simeq 0.43 \pm 0.07$ irrespective of the boundary conditions.
Therefore we can conclude that the global monopole network relaxes
into the scaling regime in the RD universe. We also show time
development of $\xi_{RD}$ in case (7) under the periodic BC in Fig.
\ref{fig:xi2}. First of all, as easily seen, the results with two
identification methods agree very well. Next, $\xi_{RD}$
asymptotically becomes a constant, $0.36\pm0.01$, which is consistent
with all the above cases with $\zeta = 10$ within the standard
deviation. Hence we can also conclude that $\zeta$ does not change the
essential result.

For the MD case, we also find that after some relaxation period, the
number of global monopoles per horizon volume becomes a constant
irrespective of cosmic time under the periodic BC expect for cases
(1), (2), and (3), in which global monopoles annihilate too much due
to the boundary effect. The trend of the boundary effect is the same
as that for the RD case. Therefore, from the results of the
largest-box simulations [case (6)], we conclude that $\xi_{MD}$
converges to a constant $\xi_{MD} \simeq 0.25 \pm 0.05$ irrespective
of the boundary conditions. Thus, we have completely confirmed that
the global monopole network goes into the scaling regime in both the
RD and MD universes.

The above results can be understood by the following simple
discussion. The evolution of the number density of global monopoles
$n(t)$ is described by the Boltzmann equation,\footnote{A similar
discussion was done in \cite{YYK} in the $2+1$ dimension.}
\bea
  \frac{dn(t)}{dt} &=& - P(t) n(t) - 3 H(t) n(t), \non \\
                   &=& - \frac{n(t)}{T(t)} - \frac{3 m n(t)}{t},
  \label{eq:Bol}
\eea
where $R(t) \propto t^{m}$, $P(t)$ is the probability per unit time
that a monopole annihilates with an antimonopole, and $T(t)$ is the
period it takes for a pair of monopoles at rest with the mean
separation $l(t)$ to pair annihilate. The mean separation $l(t)$ is
given by $l(t) \equiv R(t) r_{s} = n(t)^{-1/3}$, where $r_{s}$ is the
mean comoving separation. Since a constant attractive force works
between a pair of monopoles irrespective of the separation length, we
assume that the relative velocity between them reaches unity at once
and that they do not spiral around each other for a long period of
time. Then, the period $T(t)$ is given by the following relation,
\beq
  \int^{T+t_{0}}_{t_{0}} \frac{dt}{R(t)} = \int^{r_{s}}_{0} dr,
\eeq
where $t_{0}$ is the initial time where a pair of monopoles are at
rest. Then, the period $T(t)$ reads
\beq
  T(t) \simeq \frac{1-m}{n(t)^{1/3}}, \qquad\qquad 
     ({\rm for}~~t_{0} \ll T).
\eeq
Inserting this into the Eq. (\ref{eq:Bol}), the number density
$n(t)$ takes the following asymptotic value,
\beq
  n(t) \simeq \frac{27(1-m)^{6}}{t^{3}} \propto t^{-3}.
\eeq
From the above asymptotic form, we find that $n_{RD}(t) \simeq
0.42/t^{3}$ and $n_{MD}(t) \simeq 0.04/t^{3}$. For the RD case, $n(t)$
for the simulation and the analytic estimate agrees excellently up to
the proportional coefficient $\xi_{RD}$. The difference of the
proportional coefficient $\xi_{MD}$ between the simulation and the
analytic estimate may come from the fact that cosmic expansion is so
rapid in the matter dominated universe that our assumption that the
relative velocity between them reaches unity at once breaks down to
some extent.

We summarize our results. By directly solving equations of motion for
scalar fields in the expanding universe, we have confirmed that the
global monopole network enters the scaling regime in both the RD and
MD universes. The number density of global monopoles is given by $n(t)
= (0.43\pm0.07) / t^{3}$ for the RD case and $n(t) = (0.25\pm0.05) /
t^{3}$ for the MD case. Then, density fluctuations induced by global
monopoles are given by $\delta \rho \simeq m(t)n(t) / t^{3} \sim
7.2(5.0) \sigma^{2} / t^{2} $ for the RD (MD) era, where $m(t) \simeq
4 \pi \sigma^{2} n^{-1/3}(t) \propto t$ is the mass of a global
monopole.

The author is grateful to J. Yokoyama for useful comments.  This work
was partially supported by the Japanese Grant-in-Aid for Scientific
Research from the Ministry of Education, Culture, Sports, Science, and 
Technology.

\twocolumn[\hsize\textwidth\columnwidth\hsize\csname
@twocolumnfalse\endcsname

\begin{table}
\caption{Seven different sets of simulations for the RD case.}
\label{tab:set1}
  \begin{center}
     \begin{tabular}{cccccccc}
         Case & Lattice &
         Lattice spacing ($\delta x$) & $\zeta$ & Realization 
         & ${\rm Box~size}/H^{-1}$
         & $\xi$  & $\xi$ \\
         & number & \protect{[}unit = $t_{i}R(t)$\protect{]} &  &  & (at final time) &
         (periodic BC) & (reflective BC) \\
        \hline
        (1) & $128^3$ & $\sqrt{3}/10$ & 10 & 50 & 1 (at 125) &
        $0.28\pm0.19$ & $1.00\pm035$\\ 
        (2) & $256^3$ & $\sqrt{6}/20$ & 10 & 10 & 1 (at 250) &
        $0.31\pm0.20$ & $0.83\pm0.15$\\
        (3) & $256^3$ & $\sqrt{3}/20$ & 10 & 10 & 1 (at 125) &
        $0.35\pm0.21$ & $0.71\pm0.33$\\
        (4) & $128^3$ & $\sqrt{3}/5$  & 10 & 50 & 2 (at 125) &
        $0.37\pm0.06$ & $0.71\pm0.11$\\
        (5) & $256^3$ & $\sqrt{3}/10$ & 10 & 10 & 2 (at 125) &
        $0.36\pm0.07$ & $0.58\pm0.12$\\
        (6) & $256^3$ & $\sqrt{3}/5$  & 10 & 10 & 4 (at 125) &
        $0.36\pm0.01$ & $0.50\pm0.03$\\
        \hline
        (7) & $128^3$ & $\sqrt{6}/10$ & 5  & 50 & 1 (at 250) &
        $0.36\pm0.17$ & $0.61\pm0.21$\\
     \end{tabular}
  \end{center}
\end{table}
\begin{table}
\caption{Seven different sets of simulations for the MD case.}
\label{tab:set2}
  \begin{center}
     \begin{tabular}{cccccccc}
         Case & Lattice &
         Lattice spacing ($\delta x$) & $\zeta$ & Realization 
         & ${\rm Box~size}/H^{-1}$
         & $\xi$  & $\xi$ \\
         & number & \protect{[}unit = $t_{i}R(t)$\protect{]} &  &  & (at final time) &
         (periodic BC) & (reflective BC) \\
        \hline
        (1) & $128^3$ & $3(100)^{1/3}/256$ & 10 & 50 & 1 (at 100) &
        Disappearance & $0.75\pm0.36$\\ 
        (2) & $256^3$ & $3(200)^{1/3}/512$ & 10 & 10 & 1 (at 200) &
        Disappearance & $0.77\pm0.10$\\
        (3) & $256^3$ & $3(100)^{1/3}/512$ & 10 & 10 & 1 (at 100) &
        Disappearance & $0.82\pm0.50$\\
        (4) & $128^3$ & $3(100)^{1/3}/128$ & 10 & 50 & 2 (at 100) &
        $0.19\pm0.09$ & $0.42\pm0.10$\\
        (5) & $256^3$ & $3(100)^{1/3}/256$ & 10 & 10 & 2 (at 100) &
        $0.21\pm0.09$ & $0.41\pm0.05$\\
        (6) & $256^3$ & $3(100)^{1/3}/128$ & 10 & 10 & 4 (at 100) &
        $0.20\pm0.02$ & $0.30\pm0.02$\\
        \hline
        (7) & $128^3$ & $3(200)^{1/3}/256$ & 5  & 50 & 1 (at 200) &
        Disappearance & $0.44\pm0.03$ \\
     \end{tabular}
  \end{center}
\end{table}

]


\begin{references}

\bib{KIB}T. W. B. Kibble,
\JP{9}{1387}{76}.

\bib{TD}See, for a review, A. Vilenkin and E. P. S. Shellard, \new{\it
  Cosmic String and Other Topological Defects}\ (Cambridge University
Press, Cambridge, England, 1994).

\bib{BOOMERANG}P. de Bernardis {\it et al.}, 
\NATT{404}{955}{00}; \\
A. E. Lange {\it et al.}, 
\PRDD{63}{042001}{01}.

\bib{MAXIMA}S. Hanany {\it et al.}, 
\APJLL{545}{5}{00}; \\
A. Balbi {\it et al.}, 
\IBID{545}{L1}{00}.

\bib{TDCMB}F. R. Bouchet, P. Peter, A. Riazuelo, and M. Sakellariadou,
astro-ph/0005022;
C. R. Contaldi,
astro-ph/0005115.

\bib{Yokoyama}J. Yokoyama,
\PRL{63}{712}{89}; \PLB{212}{273}{88}.

\bib{nonG}P .G. Ferreira, J. Magueijo, and K. M. Gorski,
\APJL{503}{1}{98};
J. Pando, D. Valls-Gabaud, and L. Fang,
\PRL{81}{4568}{98};
D. Novikov, H. Feldman, and S. Shandarin,
\IJMPD{8}{291}{99}.

\bib{local}A. Albrecht and N. Turok,
\PRL{54}{1868}{85}; \PRD{40}{973}{89}; 
D. P. Bennett and F. R. Bouchet,
\PRL{60}{257}{88};
\IB{63}{2776}{89};
\PRD{41}{2408}{90};
B. Allen and E. P. S. Shellard,
\PRL{64}{119}{90}.

\bib{VHS}G. R. Vincent, M. Hindmarsh, and M. Sakellariadou,
\PRD{56}{637}{97};
G. R. Vincent, N. D. Antunes, and M. Hindmarsh,
\PRL{80}{2277}{98}.

\bib{global}M. Yamaguchi, M. Kawasaki, and J. Yokoyama,
\PRL{82}{4578}{99}.

\bib{global2}
M. Yamaguchi,
\PRD{60}{103511}{99};
M. Yamaguchi, M. Kawasaki, and J. Yokoyama,
\IBIDD{61}{061301}{00}.

\bib{axion}R. L. Davis, 
\PRD{32}{3172}{85}; \PLB{180}{225}{86};
R. L. Davis and E. P. S. Shellard, 
\NPB{324}{167}{89};
A. Dabholkar and J. M. Quashnock, 
\IBID{333}{815}{90};
R. A. Battye and E. P. S. Shellard, 
\IBID{423}{260}{94}; \PRL{73}{2954}{94}.

\bib{axion2}D. Harari and P. Sikivie, 
\PLB{195}{361}{87};
C. Hagmann and P. Sikivie, 
\NPB{363}{247}{91};
C. Hagmann, S. Chang, and P. Sikivie, 
\PRDD{63}{125015}{01}.

\bib{inflation}See, for example, A. D. Linde, {\it Particle Physics
and Inflationary Cosmology} (Harwood, Chur, Switzerland, 1990).

\bib{LP}P. Langacker and S.-Y. Pi,
\PRL{45}{1}{80}.

\bib{DLV}G. Dvali, H. Liu, and T. Vachaspati,
\PRL{80}{2281}{98}.

\bib{BV}M. Barriola and A. Vilenkin,
\PRL{63}{341}{89}.

\bib{BR}D. P. Bennett and S. H. Rhie,
\PRL{65}{1709}{90}.

\bib{PST}U. Pen, D. N. Spergel, and N. Turok,
\PRD{49}{692}{94}.

\bib{DKM}M. Kunz and R. Durrer,
\PRD{55}{R4516}{97},
R. Durrer, M. Kunz, and A. Melchiorri,
\IBID{59}{123005}{99}.

\bib{YYK}M. Yamaguchi, J. Yokoyama, and M. Kawasaki,
\PTP{100}{535}{98}.


\newpage

\end{references}
\end{document}